\documentclass[conference]{IEEEtran}
\IEEEoverridecommandlockouts
\usepackage{cite}
\usepackage{amsmath,amssymb,amsfonts}
\usepackage{algorithm}
\usepackage{algpseudocode}
\usepackage{graphicx}
\usepackage{textcomp}
\usepackage{booktabs}
\usepackage{xcolor}
\usepackage{url}
\usepackage{hyperref}
\def\BibTeX{{\rm B\kern-.05em{\sc i\kern-.025em b}\kern-.08em
    T\kern-.1667em\lower.7ex\hbox{E}\kern-.125emX}}

\usepackage[most]{tcolorbox}

\tcbset{
  rqbox/.style={
    colback=gray!5,
    colframe=black!40,
    boxrule=0.5pt,
    arc=1mm,
    left=4pt,right=4pt,top=4pt,bottom=4pt,
  },
}

\begin{document}

\title{QBugLM: An Agentic Benchmarking Framework for LLM-based Quantum Software Debugging}

% \title{QBugLM: An Agentic Benchmarking Framework for LLM-based Quantum Software Bug Detection and Repair}

% \title{QBugLM: Benchmarking LLM-based Bug Detection and Repair for Quantum Software}

\author{
    \IEEEauthorblockN{An B. B. Pham$^{\dagger}$\textsuperscript{1}, Hoa T. Nguyen$^{\dagger}$\textsuperscript{2}, and Muhammad Usman\textsuperscript{2,3}}
    \IEEEauthorblockA{\textsuperscript{1}\textit{The University of Melbourne, Parkville, 3052, Victoria, Australia}}
    \IEEEauthorblockA{\textsuperscript{2}\textit{Data61, CSIRO, Clayton, 3168, Victoria, Australia}} 
    \IEEEauthorblockA{\textsuperscript{3}\textit{School of Physics, The University of Melbourne, Parkville, 3052, Victoria, Australia}}
    a.pham@unimelb.edu.au, \{hoa.nguyen, muhammad.usman\}@csiro.au
    \thanks{$^{\dagger}$These authors contributed equally.}
}

% \author{
% Anonymous Author(s)$^\dagger$%
% \thanks{$\dagger$ Paper submitted for double-blind review.}%
% }

\maketitle

\begin{abstract}
Quantum software bugs often yield silent, incorrect outputs rather than explicit errors, making them particularly difficult to detect and repair with conventional techniques. Although large language models (LLMs) have shown strong performance on classical software engineering tasks, their ability to debug quantum code remains largely unexplored. To bridge this gap, we propose QBugLM, a multi-agent framework that automates the quantum software debugging pipeline, from taxonomy-driven bug injection to LLM-based detection and repair, and finally to simulation-based validation, for framework-agnostic OpenQASM 3.0 programs. We further conduct a comprehensive case study using QBugLM to benchmark two LLMs, Claude 4.6 Sonnet and Qwen3 Coder Next, across different prompting strategies, bug categories, and quantum programs. Our results show that iterative feedback is critical, as a single retry raises Pass@1 from below 25\% to above 80\%. Moreover, simpler structured prompting can even outperform Chain-of-Thought and ReAct for reasoning-capable models under fixed-resource constraints. Our work takes initial steps toward benchmarking LLM capabilities for debugging quantum programs and offers practical insights to support future efforts in automated quantum software repair.
\end{abstract}


\section{Introduction}
Quantum software engineering poses challenges that are qualitatively distinct from those of classical software development. The probabilistic nature of quantum computation and the absence of mature debugging toolchains make bugs in quantum programs exceptionally difficult to isolate and diagnose \cite{leite_ramalho_testing_2025}. In contrast to classical program bugs, which typically manifest as explicit exceptions or system crashes, quantum-specific bugs often produce silent, incorrect outputs, substantially undermining the effectiveness of conventional debugging techniques \cite{paltenghi_bugs_2022, zhao_bugs4q_2023}. As quantum hardware matures and quantum software toolkits such as Qiskit, Cirq, and PennyLane scale up and become more complex, the demand for automated support mechanisms in quantum software engineering \cite{sarkar_automated_2024} continues to grow.

In the meantime, large language models (LLMs) have shown strong performance on common software engineering tasks \cite{fan_large_2023, he_llm-based_2025}, including code generation, fault localisation, and automated program repair. Early investigations assessed the capacity of general-purpose LLMs to generate quantum code. For instance, Henderson et al. \cite{henderson_programming_2025} found that models such as GPT-4 performed well on generating simple discrete-variable quantum circuits in Qiskit, but struggle with continuous-variable tasks in Xanadu's Strawberry Fields, frequently violating hardware constraints and producing logically incorrect circuits. To address these shortcomings, several domain-adapted models have been proposed. The Qiskit Code Assistant \cite{dupuis_qiskitcodeassistant_2024}, built upon IBM Granite model and fine-tuned on curated Qiskit scripts and notebooks, substantially outperforms general-purpose models in generating functionally correct Qiskit code. Similarly, PennyCoder \cite{basit_pennycoder_2025} applies Low-Rank Adaptation (LoRA) to fine-tune LLaMA models on the PennyLang dataset, enabling efficient on-device quantum code generation for quantum machine learning and quantum reinforcement learning tasks using PennyLane. AGENT-Q \cite{jern_agent-q_2025} fine-tuned on over 14,000 optimised quantum circuits, excels at generating parameterised circuits for algorithms such as QAOA and VQE, producing initial parameters closer to optimal than random initialisation.

Recent studies have also benchmarked LLM performance on quantum code generation tasks. For example, QuanBench \cite{guo_quanbench_2025} evaluates nine LLMs across 44 quantum programming tasks spanning algorithms, state preparation, gate decomposition, and quantum machine learning. It reports overall accuracy below 40\% and highlights recurring failure modes, including circuit construction errors and flawed algorithm logic. QHackBench \cite{basit_qhackbench_2025} evaluates LLMs on PennyLane-based quantum programming tasks drawn from real-world hackathon challenges, with the best-performing model achieving approximately 47 to 49\% accuracy. Qiskit HumanEval \cite{vishwakarma_qiskit_2024} assesses the functional correctness of LLM-generated Qiskit code, with an emphasis on API usage and syntactic correctness.

% While LLMs have demonstrated strong performance in classical code generation, quantum programming introduces distinct challenges, including probabilistic execution semantics and specialised low-level compilation targets, such as OpenQASM \cite{guo_quanbench_2025, fu_qagent_2025}. These challenges have motivated a growing body of studies spanning model specialisation, agentic frameworks, benchmark design, and automated program repair. 
% Rigorous evaluation of LLM-generated quantum code requires benchmarks that go beyond syntactic correctness to assess quantum semantics. Qiskit HumanEval \cite{vishwakarma_qiskit_2024} adapts the classical HumanEval benchmark to over 100 quantum tasks, testing functional correctness within the Qiskit framework. QuanBench \cite{guo_quanbench_2025} evaluates nine LLMs across 44 complex tasks spanning algorithm implementation, state preparation, gate decomposition, and quantum machine learning, measuring both functional correctness (Pass@K) and process fidelity, a metric that quantifies the similarity between the unitary matrix of the generated circuit and that of the canonical solution. QHackBench \cite{basit_qhackbench_2025} targets the PennyLane framework using historical hackathon challenges and focuses on quantum machine learning and hybrid quantum-classical scenarios.

\begin{table*}[htbp]
\caption{Taxonomy of Common Bugs in Quantum Software}
\label{tab:bug_taxonomy}
\centering
\renewcommand{\arraystretch}{1.15}
\begin{tabular}{@{}p{3.9cm}p{6.9cm}p{5.8cm}@{}}
\toprule
\textbf{Category} & \textbf{Example} & \textbf{OpenQASM3 Example} \\
\midrule
C1: Deprecated Syntax Errors & Use deprecated OpenQASM 2.0 syntax in OpenQASM 3.0 program or syntax errors & \texttt{include "stdgates.inc";} $\to$ \texttt{include "qelib1.inc";} \\
\midrule
C2: Structural Errors in Circuit Construction & Violate circuit construction rules, such as assigning identical indices to control and target qubits in CNOT gates & \texttt{cx q[0],q[1]} $\to$ \texttt{cx q[0],q[0]} \\
\midrule
C3: Gate Overuse/Redundancy  & Duplicate a self-inverse quantum gate (e.g., X, Y, Z, H, CX) & \texttt{h q[0]} $\to$ \texttt{h q[0]; h q[0]; h q[0]} \\
\midrule
C4: Semantic Deviation & Substitute a gate with an incorrect gate of equal arity & \texttt{rx(0.5) q[0]} $\to$ \texttt{ry(0.5) q[0]} \\
\cmidrule(l){2-3}
 & Insert a spurious gate operation & --- $\to$ \texttt{x q[2]} \\
\cmidrule(l){2-3}
 & Delete a required gate operation & \texttt{h q[1]} $\to$ (removed) \\
\cmidrule(l){2-3}
 & Change the phase value of a rotation gate & \texttt{rz(pi/4) q[0]} $\to$ \texttt{rz(-pi/4) q[0]} \\
\cmidrule(l){2-3}
 & Duplicate a measurement statement at an incorrect location & \texttt{measure q[0]} (duplicated) \\
\cmidrule(l){2-3}
 & Remove a required measurement statement& \texttt{measure q[2]} $\to$ (removed) \\
\bottomrule
\end{tabular}
\end{table*}

However, a critical gap remains in the use of LLMs for debugging quantum software. Although LLM-based code generation has attracted growing attention, the capacity of LLMs to detect and repair bugs in existing quantum programs, especially bugs in LLM-generated quantum code, has not yet been systematically investigated. Furthermore, several existing studies focus on specific software development kits (SDKs), such as Qiskit \cite{campbell_enhancing_2025, yoshida_leveraging_2026}, tightly coupling evaluation to framework-specific code rather than to the underlying logical quantum circuits. Consequently, LLMs' ability to debug low-level languages such as OpenQASM \cite{Cross2022OpenQASM3:Language} in a quantum SDK-agnostic setting remains underexplored. This capability is equally crucial for agentic quantum software development workflows, in which an LLM agent must not only synthesise code but also iteratively diagnose failures and apply corrective edits until the resulting quantum program satisfies both syntactic and semantic correctness criteria. We address these gaps by proposing QBugLM, a novel multi-agent benchmarking framework that systematically evaluates the detection and repair capabilities of LLMs on common quantum software bugs.

The main contributions of our paper are as follows:
\begin{itemize}
    \item We propose QBugLM, a multi-agent benchmarking framework that automates the quantum software debugging pipeline, facilitating LLM-based bug detection and repair, and simulation-based validation, operating on OpenQASM 3.0 quantum SDK-agnostic programs.
    \item Within QBugLM, we also develop QBugGen, a mutation toolkit that systematically injects bugs, drawn from common quantum bug taxonomy, into valid quantum circuits, yielding a controlled evaluation dataset with ground-truth annotations for reproducible benchmarking studies.
    \item We conduct a comprehensive case study benchmarking Claude 4.6 Sonnet and Qwen3 Coder Next across different prompting strategies, bug categories, and quantum circuits, yielding three key insights, including (1) a single retry raises Pass@1 from below 25\% to above 80\%, establishing iterative feedback as the dominant factor in repair success; (2) simpler structured prompting can even outperform Chain-of-Thought and ReAct for reasoning-capable models under fixed resource constraints; and (3) the open-source LLM achieves comparable accuracy at significant lower cost on most bug types in the case study.
\end{itemize}

The rest of the paper is organised as follows. Section II introduces the proposed QBugLM framework for LLM-based quantum software debugging and a taxonomy of common bugs in quantum software. Section III provides a comprehensive case study of our framework and discusses the implications of our findings. Section IV reviews related work and highlights the research gaps our study addresses. Finally, Section V concludes the paper by summarising our contributions and outlining directions for future work.

\begin{figure*}[htbp]
\centering
\includegraphics[width=6.3in]{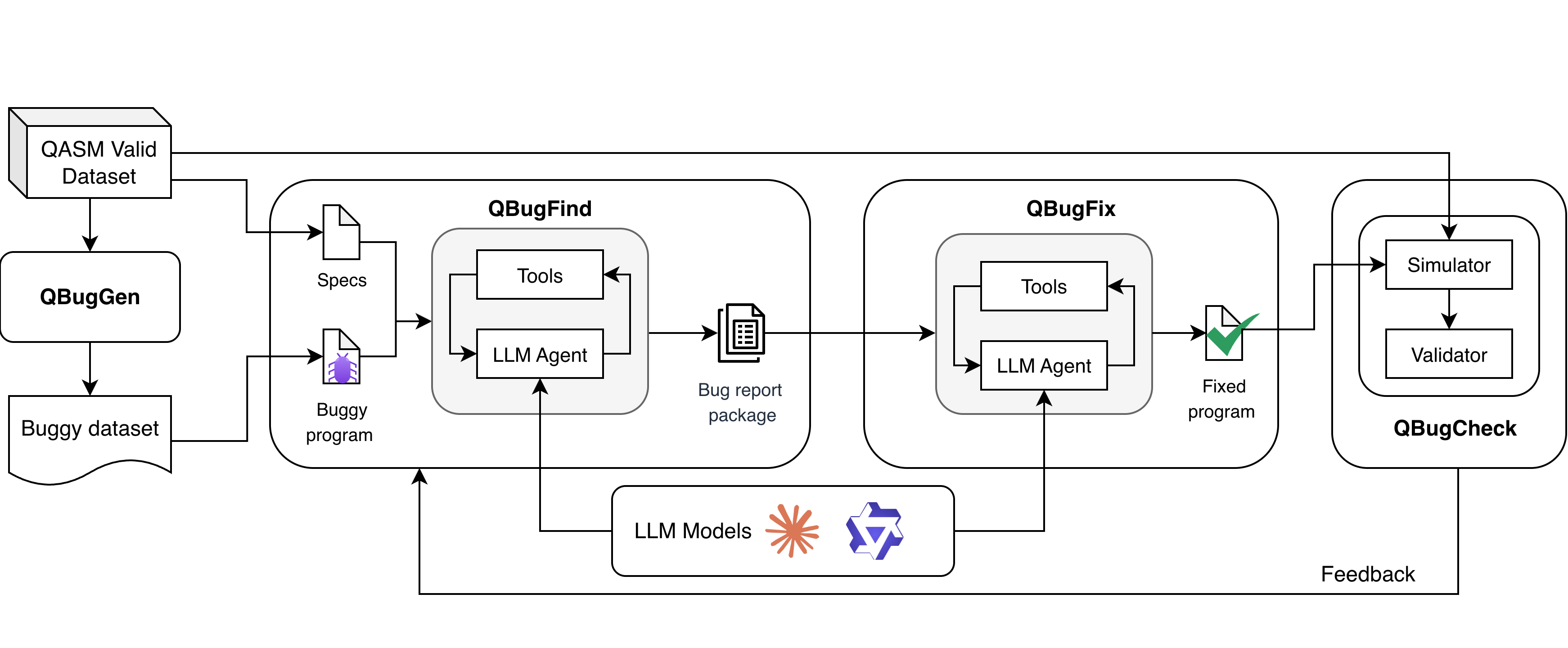}
\caption{Overview of the QBugLM framework}
\label{fig_overview}
\end{figure*}

% Furthermore, we complement the framework with a comprehensive case study benchmarking two state-of-the-art models (Claude 4.6 Sonnet and Qwen3 Coder Next) across different prompting strategies, bug categories, and quantum circuits. 
% This study, to our knowledge, provides one of the first empirical comparisons of proprietary and open-source frontier LLMs for multi-agent quantum software debugging, examining both repair accuracy and the interplay among prompting complexity, iterative feedback, program characteristics, and operational cost.

\section{QBugLM Framework}
As illustrated in Figure \ref{fig_overview}, our proposed QBugLM framework comprises four main components that collectively form an end-to-end pipeline for evaluating LLM-based quantum software debugging: 
% The pipeline proceeds sequentially: 

% (1) \textit{QBugGen} generates mutated quantum programs from a validated source dataset; (2) \textit{QBugFind} employs an LLM agent to detect and localise bugs in each mutant; (3) \textit{QBugFix} employs a second LLM agent to produce a corrected program based on the bug report; and (4) \textit{QBugCheck} validates the repaired program against the original ground-truth circuit. 

% Each component is described in detail below.

\subsubsection{QBugGen} takes as input a corpus of syntactically and semantically valid OpenQASM  3.0 programs sourced from an well-known quantum circuit benchmark suite (e.g., MQT Bench \cite{quetschlich_mqtbenchbenchmarking_2023}). Building on the literature on quantum bugs \cite{paltenghi_bugs_2022, zhao_bugs4q_2023, fortunato_qmutpy_2022} and on failure-mode analyses of LLM-generated quantum code reported in prior benchmarking studies \cite{ guo_quanbench_2025}, we consolidate common quantum software bugs into a four-category taxonomy that facilitates QBugGen, with representative examples in Table \ref{tab:bug_taxonomy}. For each valid program, QBugGen systematically injects a single, well-defined bug drawn from the taxonomy presented. Each mutation is recorded as a structured specification that captures the mutation type, the affected source line, and the applied transformation. 

\subsubsection{QBugFind} takes a buggy quantum program from QBugGen and invokes an LLM agent to generate a structured bug report. The agent is given the buggy source code, program specifications, and a prompt derived from a configurable strategy. It identifies the bug's location (i.e., the affected line) and classifies it according to the bug taxonomy. QBugFind outputs a bug-report package containing the suspected fault location, predicted bug category, the original specifications and buggy program, which is then passed to QBugFix.

\subsubsection{QBugFix} receives the original buggy program alongside the bug report generated by QBugFind and delegates the repair task to a second LLM agent to produce a corrected version of the quantum program that resolves the identified bugs while preserving the intended algorithmic semantics. The agent is unconstrained in the repair operations it may apply, including substitution, insertion, removal of gates, instruction reordering, modification of rotation parameters, and adjustment of qubit indices. Separating detection and repair into distinct agents can enable independent evaluation of each capability and systematic study of how detection quality affects repair outcomes. 

\subsubsection{QBugCheck} operates as a deterministic validator that compares the behaviour of the LLM-fixed program against the original ground-truth circuit from the source dataset. We determine whether a repaired (fixed) program is functionally equivalent to the ground-truth (reference) program by comparing total variation distance $\delta$ in measurement distributions. 

\begin{equation}
    \delta(P_{ref}, P_{fixed}) = \frac{1}{2} \sum_k |P_{ref}(k) - P_{fixed}(k)|
\end{equation}
where $P_{ref}$ and $P_{fixed}$ are probability distributions of the reference program and fixed program, respectively, and $k$ indexes each possible measurement outcome. Both the reference and fixed programs are executed on a noiseless quantum simulator to eliminate hardware noise, so any distributional divergence reflects logical differences between the programs. The fix is accepted if the total variation distance across all outcome probabilities satisfies $\delta(P_{ref}, P_{fixed}) \leq \varepsilon_\delta$, where $\varepsilon_\delta$ is the accuracy threshold. We additionally compare gate counts at the same circuit transpilation optimisation level, and the structural check passes when $|G_{ref} - G_{fixed}| \leq \varepsilon_g$, where $\varepsilon_g$ is the gate count tolerance, $G_{ref}$ and $G_{fixed}$ denote the transpiled gate counts of the reference and fixed programs, respectively.

The end-to-end debugging pipeline of the proposed framework is illustrated in Algorithm~\ref{alg:qbuglm}. The algorithm accepts as input a set of valid quantum programs $\mathcal{P}$, a collection of mutation operators $\mathcal{M}$, two LLM agents $\mathcal{A}_{\text{find}}$ and $\mathcal{A}_{\text{fix}}$, a prompting strategy $\sigma$, a maximum number of attempts $K$, a distribution tolerance $\varepsilon_\delta$, and a gate count tolerance $\varepsilon_g$. The output is a result set $\mathcal{R}$ that records, for every mutant, the final bug report, the final candidate fix, the repair outcome, the number of attempts consumed, and the validation verdict. 

\begin{algorithm}[htbp]
  \caption{QBugLM End-to-End Workflow}
  \label{alg:qbuglm}
  \begin{algorithmic}[1]
  \Require Valid program set $\mathcal{P}$, mutation operators $\mathcal{M}$, LLM agents $\mathcal{A}_{\text{find}}$ and $\mathcal{A}_{\text{fix}}$, prompting
  strategy $\sigma$, max attempts $K$, tolerance thresholds $\varepsilon_\delta$ and $\varepsilon_g$
  \Ensure Results set $\mathcal{R}$ with per-instance repair
  \State $\mathcal{R} \gets \emptyset$
  \For{each valid program $p \in \mathcal{P}$}
      \State \textbf{Stage 1: QBugGen}
      \State $\mathcal{B}_p \gets \textsc{QBugGen}(p, \mathcal{M})$ \Comment{Generate mutants}
      \State $s_p \gets \textsc{Spec}(p)$ \Comment{Program specification}
      \For{each $p_{\text{bug}} \in \mathcal{B}_p$}
          \State $\textit{history} \gets \emptyset$; \enspace $\textit{repaired} \gets \textsc{False}$
          \For{$k \gets 1$ \textbf{to} $K$}
              \State \textbf{Stage 2: QBugFind}
              \State $r_{\text{bug}} \gets \mathcal{A}_{\text{find}}(p_{\text{bug}},\, s_p,\, \sigma,\, \textit{history})$ 
              \State \textbf{Stage 3: QBugFix}
              \State $p_{\text{fix}} \gets \mathcal{A}_{\text{fix}}(p_{\text{bug}},\, r_{\text{bug}},\, s_p,\, \sigma,\, \textit{history})$ 
              \State \textbf{Stage 4: QBugCheck}
              \State $v \gets \textsc{QBugCheck}(p,\, p_{\text{fixed}},\, \varepsilon_\delta,\, \varepsilon_g)$ 
              \If{$v.\textit{pass} = \textsc{True}$}
                  \State $\textit{repaired} \gets \textsc{True}$
                  \State \textbf{break}
              \EndIf
              \State $\textit{history} \gets \textit{history} \cup \{(r_{\text{bug}},\, p_{\text{fixed}},\, v)\}$ 
          \EndFor
          \State $\mathcal{R} \gets \mathcal{R} \cup \{(p_{\text{bug}},\, r_{\text{bug}},\, p_{\text{fixed}},\, k,\, v)\}$
      \EndFor
  \EndFor
  \State \Return $\mathcal{R}$
  \end{algorithmic}
  \end{algorithm}

\section{Case Study: QBugLM Benchmarking}
To demonstrate the practical utility of the proposed framework, we conduct a comprehensive case study using QBugLM to evaluate the performance of state-of-the-art LLMs in quantum software bug detection and repair. 

\subsection{Research Questions}
We evaluate the capabilities of LLMs in quantum software bug detection and repair using our QBugLM framework through the following research questions (RQs):

\begin{itemize}
    \item \textbf{RQ1}: How does the choice of prompting strategy affect LLM performance in quantum software debugging? 
    
    \item \textbf{RQ2}: What is the current capability of LLMs in detecting and repairing different types of bugs in quantum programs under varying retry constraints? 
    
    \item \textbf{RQ3}: How cost-efficient are different LLMs for quantum software debugging? 
\end{itemize}

\subsection{Experiment Setup}
We implement the proposed framework using Strands Agents\footnote{https://strandsagents.com (accessed January 2026)} as the multi-agent orchestration backbone. The pipeline is deployed on a single workstation, and all quantum circuit simulations are run via a noiseless quantum simulator using Qiskit Aer\footnote{https://github.com/Qiskit/qiskit-aer (accessed January 2026)}. For each experimental configuration, we set the maximum number of attempts to $K = 3$ (i.e., one initial attempt plus two retries). All simulation-based validations use $N = 1{,}024$ shots, with a distribution tolerance of $\varepsilon_\delta = 0.05$ and a gate count tolerance of $\varepsilon_g = 0$. 

\subsubsection{Quantum programs}
We construct the ground-truth corpus from MQT Bench \cite{quetschlich_mqtbenchbenchmarking_2023}, selecting five representative 5-qubit OpenQASM 3.0 circuits, including Deutsch-Jozsa's algorithm (dj), Grover's algorithm (grover), Bernstein-Vazirani's algorithm (bv), GHZ state (ghz) and W state (wstate).

\subsubsection{LLMs Selection}
To assess the influence of model scale and provenance on debugging performance, we evaluate two representative LLMs: 1) a proprietary model with Claude Sonnet 4.6 (Anthropic) and 2) an open-source, code-specialised model with Qwen3 Coder Next (Alibaba). For simplicity, we also refer to these as Sonnet 4.6 and Qwen3, respectively.

\subsubsection{Prompting Strategies}
We evaluate three main common prompting strategies to assess the impact of in-context guidance on detection and repair, including: 1) \textit{Structured} (general), 2) \textit{Chain-of-Thought (CoT)} \cite{wei_chain--thought_2023}, and 3) \textit{ReAct} \cite{yao_react_2023}. 

\textbf{Code Availability:} All source code and prompting examples of QBugLM are available at \href{https://github.com/qachub/qbuglm}{github.com/qachub/qbuglm}.

\subsection{Evaluation Metrics}
We evaluate the framework along two main dimensions: correctness of the debugging and efficiency of the pipeline.

\textbf{Correctness metrics.} 
To assess cross-run and overall reliability, we adopt the unbiased Pass@$k$ estimator following prior studies \cite{kulal_spoc_2019, chen_evaluating_2021, guo_quanbench_2025}. For each mutant, the pipeline is executed $n$ times independently, yielding $c$ correct fixes. The Pass@$k$ metric estimates the probability that at least one of $k$ randomly chosen samples is correct:
\begin{equation}
    \text{Pass@}k = 1 - \dfrac{\binom{n - c}{k}}{\binom{n}{k}}
\end{equation}
where $n$ is the total number of independent runs per task, $c$ is the number of runs that produce a correct fix (as verified by QBugCheck), and $k \leq n$ is the number of samples drawn.

\textbf{Efficiency metrics.} To assess the operational cost of LLM-based debugging, we measure: 1) \textit{Token consumption.} The total number of input and output tokens consumed by the QBugFind and QBugFix agents per mutant, reported in aggregate; 2) \textit{Wall-clock time.} The elapsed time per mutant from pipeline start to final validation verdict; and 3) \textit{Monetary cost.} The estimated average API cost per mutant, according to AWS Bedrock pricing (as of March 2026)\footnote{https://aws.amazon.com/bedrock/pricing/ (accessed March 2026)}.

\subsection{Experiment Results}
\subsubsection{RQ1} \textit{How does the choice of prompting strategy affect LLM performance in quantum software debugging?}

\begin{figure}[htbp]
    \centering
    \includegraphics[width=0.7\columnwidth]{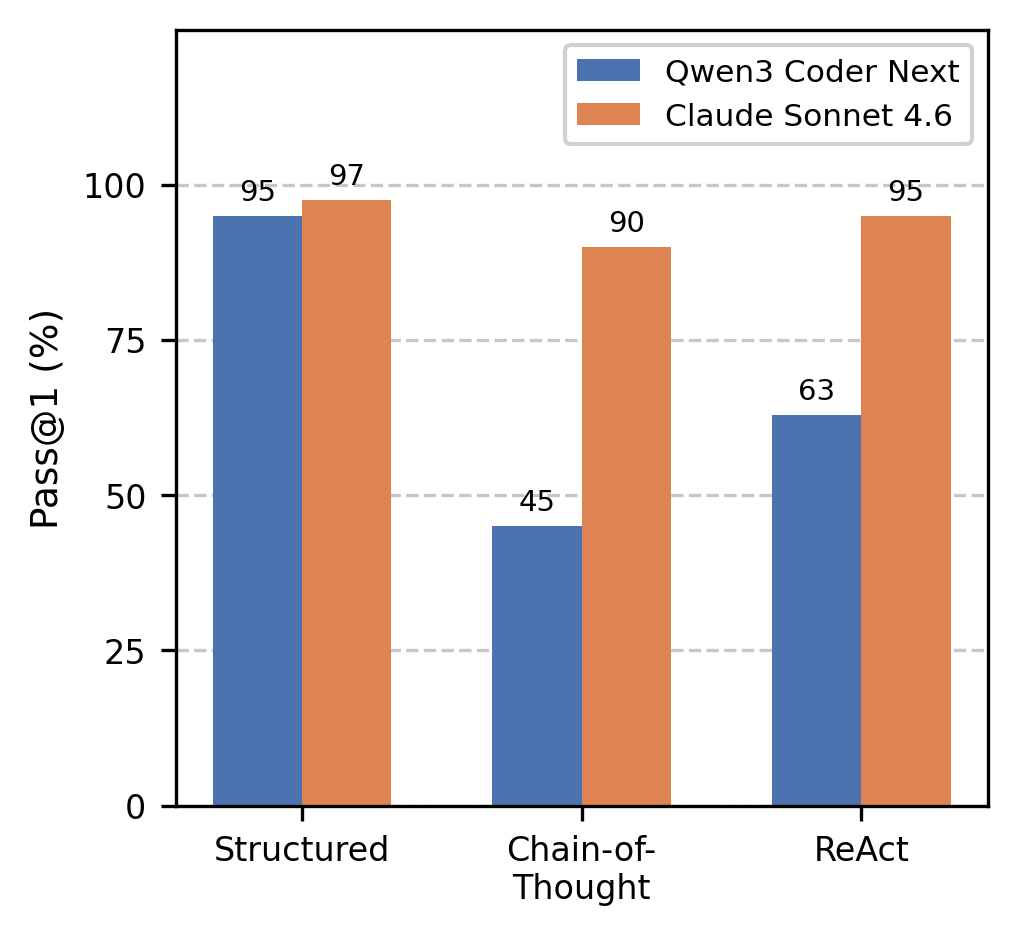}
    \caption{Pass@1 of Qwen3 Coder Next and Claude Sonnet 4.6 across three prompting strategies on the Bernstein-Vazirani (BV) circuit. Structured prompting achieves the highest Pass@1 for both models.}
    \label{fig_rq1_result}
\end{figure}

Figure~\ref{fig_rq1_result} presents the Pass@1 of both models across three prompting strategies on the Bernstein-Vazirani circuit. Structured prompting achieves the highest performance, with Claude Sonnet 4.6 reaching 97\% and Qwen3 Coder Next reaching 95\%. Both models perform worse under CoT and ReAct, with CoT reducing success rates to 90\% and 45\%, respectively, and ReAct to 95\% and 63\%, respectively. This result is contrary to the expectation that explicit reasoning scaffolds improve LLM performance. Both models are natively reasoning-capable and can perform internal deliberation without explicit instruction \cite{hu_asking_2026}. 

\begin{tcolorbox}[rqbox,title={Answer to RQ1}]
Prompting strategy has a substantial impact on LLM-based quantum debugging, as structured prompting consistently outperforms CoT and ReAct across both models, challenging the assumption that explicit reasoning scaffolds help capable LLMs and showing that, in resource-constrained pipelines, such scaffolds can even degrade overall debugging performance.
\end{tcolorbox}

\begin{figure*}[htbp]
    \centering
    
    \begin{minipage}[t]{0.32\textwidth}
        \centering
        \includegraphics[width=\linewidth]{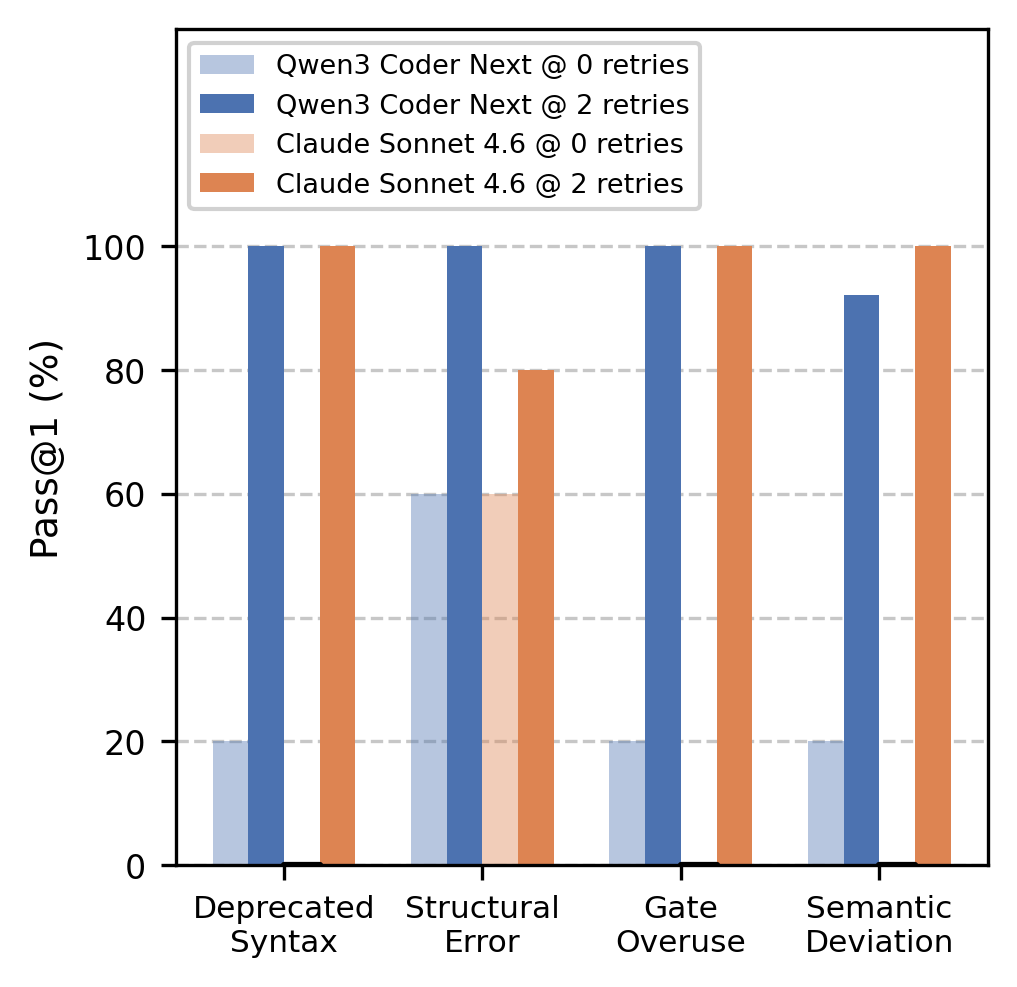}
        \caption{
        Pass@1 per bug category at zero and two retries for both LLMs on the BV circuit with structured prompting.
        } 
        \label{fig_rq2_category}
    \end{minipage}
    \hfill
    \begin{minipage}[t]{0.32\textwidth}
        \centering
        \includegraphics[width=\linewidth]{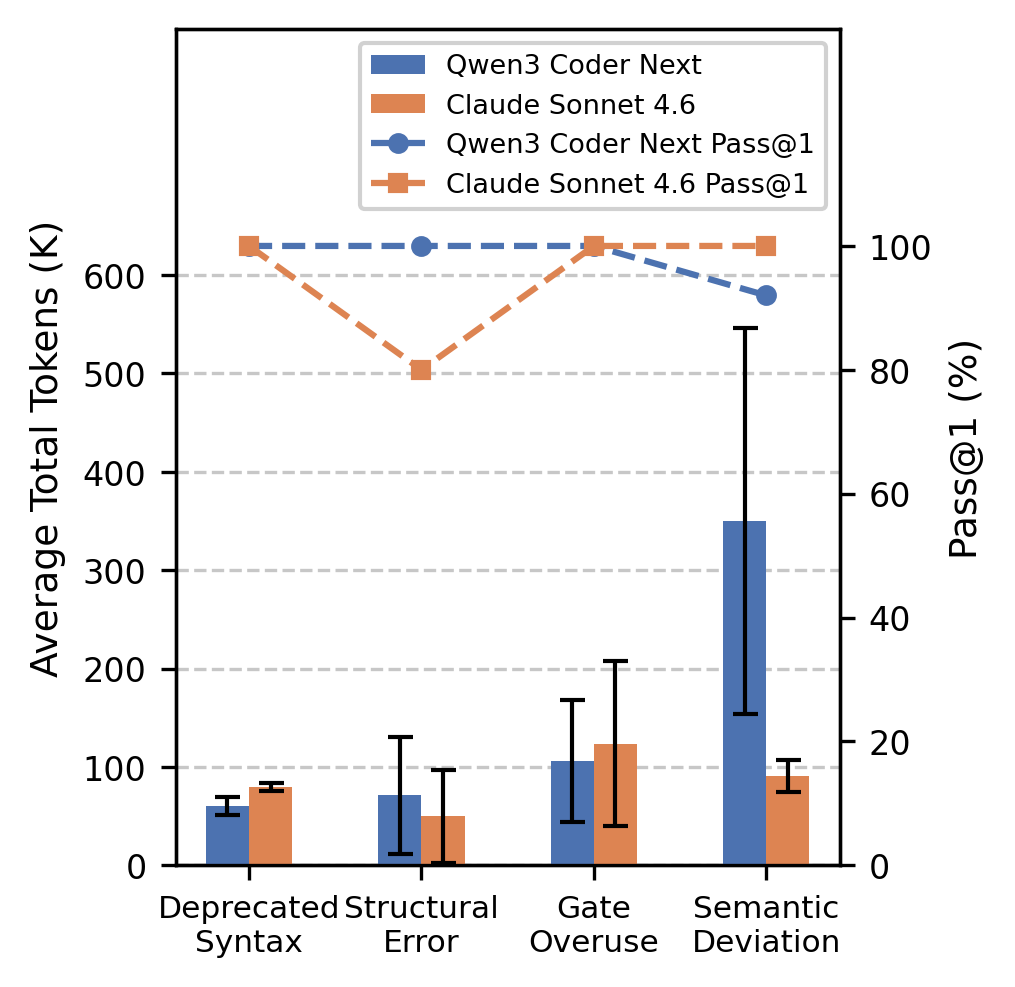}
        \caption{
        Average total token consumption (bars) and Pass@1 (dashed lines) per mutant per bug category at two retries for both LLMs. 
        }
        \label{fig_rq4_tokens}
    \end{minipage}
    \hfill
    \begin{minipage}[t]{0.32\textwidth}
        \centering
        \includegraphics[width=\linewidth]{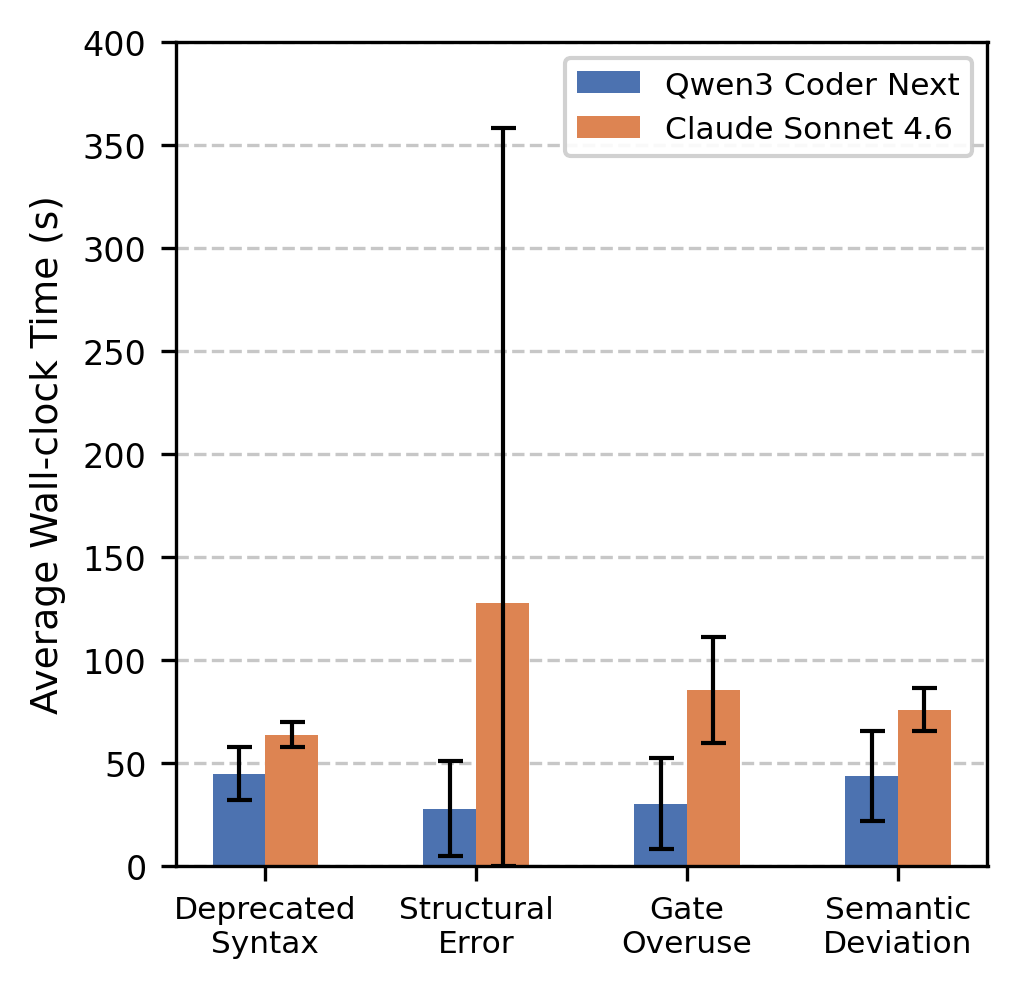}
        \caption{
        Average wall-clock time per mutant per bug category at two retries for both LLMs.
        }
        \label{fig_rq4_time}
    \end{minipage}

\end{figure*}

\subsubsection{RQ2} \textit{What is the current capability of LLMs in detecting and repairing different types of bugs in quantum programs under varying retry constraints?}

Figure~\ref{fig_rq2_category} shows Pass@1 per bug category at zero and two retries. Without retries, the Qwen3 model achieves 20\% across semantic deviation, deprecated syntax, and gate overuse, while Sonnet 4.6 fails entirely on those three bug categories. Both models achieve 60\% structural error, making it the only category in which single-shot debugging is partially reliable. After two retries, both models recover strongly across all categories, with Pass@5 reaching 100\% in all cases. Sonnet 4.6 reaches 100\% on semantic deviation, deprecated syntax, and gate overuse, but improves only to 80\% on structural error. Qwen3 reaches 100\% on these first three bug categories, but achieves only 92\% on semantic deviation. Each model, therefore, retains one unresolved weakness after two retries: Sonnet 4.6 for structural error and Qwen3 for semantic deviation. Both weaknesses persist despite full recovery across all other categories, suggesting that they represent model-specific limitations rather than artefacts of an insufficient retry budget.

\begin{tcolorbox}[rqbox, title={Answer to RQ2}]
Iterative refinement with accumulated context is essential for LLM‑based quantum program debugging, as the first retry yields the largest accuracy gain, while subsequent retries exhibit diminishing returns. A two-retry budget offers a favourable trade-off between accuracy and cost.
\end{tcolorbox}

\subsubsection{RQ3} \textit{How cost-efficient are different LLMs for quantum software debugging?}

Figure~\ref{fig_rq4_tokens} presents the average total token consumption and Pass@1 per bug category at two retries. Across structural error, deprecated syntax, and gate overuse, Qwen3 achieves equal or better Pass@1 than Sonnet 4.6 while consuming fewer tokens. In terms of monetary costs, Qwen3 Coder Next costs \$0.042, \$0.036, and \$0.061 per mutant on these three categories, respectively, compared to \$0.202, \$0.327, and \$0.496 for Sonnet 4.6, representing cost reductions by factors of 4.8, 9.1, and 8.1. Figure~\ref{fig_rq4_time} further shows that Qwen3 is between 1.5 and 4.6 times faster across those same categories, with the most notable gap on structural error, where Claude Sonnet 4.6 takes 127.6 seconds on average compared to 28 seconds for Qwen3.
Semantic deviation is the exception across all three dimensions. Qwen3 consumes approximately 350,000 tokens on average for this category compared to 91,000 for Sonnet 4.6, incurring a cost of \$0.198 versus \$0.369. Despite this disproportionate token consumption, Qwen3 still achieves only 92\% Pass@1 compared to 100\% for Sonnet 4.6. This is the only category in which Sonnet 4.6's higher cost correlates with a meaningful gain in accuracy. Structural error shows the opposite trend: Sonnet 4.6 reaches only 80\% Pass@1, is 4.6 times slower, and costs 4.8 times more than Qwen3, which attains 100\% Pass@1. 

\begin{tcolorbox}[rqbox, title={Answer to RQ3}]
Overall, Qwen3 is substantially more cost-efficient than Sonnet 4.6 across most bug categories except semantic deviations, delivering equal or better Pass@1 at 4 to 9 times lower cost and 1.5 to 4.6 times faster wall-clock time. A cost-efficient deployment strategy should therefore be bug-type-aware, as Qwen3 may be preferable for structural errors, deprecated syntax, and gate overuse, while Sonnet 4.6 is more reliable for semantic deviations in this case.
\end{tcolorbox}

\subsection{Threat to Validity}
We identify two main threats to the validity of our case study. 
First, we focus on semantic bugs in quantum programs, which create a class imbalance because six mutation strategies produce semantic deviations. This makes category-level results for the first three bug categories (C1-C3) highly sensitive to single failures. We mitigate this by reporting results both per category and in aggregate. Additionally, we perform five independent runs per configuration and use the unbiased Pass@k estimator to account for sampling variance.
Second, we count infrastructure failures, such as network timeouts and per-agent context-window exhaustion, as unsuccessful fixes rather than excluding them; although this may understate raw LLM capability, it reflects realistic deployment conditions and yields a more conservative estimate. 

\section{Related Work}
Prior work on quantum software bug detection and repair spans several areas, including benchmark datasets, static and dynamic analysis, formal verification, and automated program repair (APR). First, evaluating quantum debugging tools requires curated datasets of reproducible bugs. QBugs \cite{campos_qbugs_2021} offers a complementary set of reproducible algorithmic bugs. Bugs4Q \cite{zhao_bugs4q_2023} provides a manually validated collection of Qiskit bugs sourced from GitHub and Stack Exchange, pairing each buggy program with its fix and a reproducing test case. Both benchmarks capture naturally occurring defects but remain framework-specific (primarily Qiskit), lack systematic coverage of common quantum bug taxonomy, and are not designed to evaluate LLM-based debugging in controlled settings. 

Regarding bug detection, several static analysis tools analyse quantum programs without execution. For example, QChecker \cite{zhao_qchecker_2023} uses Abstract Syntax Trees to identify recurring patterns, including incorrect gate usage, measurement anomalies, and deprecated API calls. ScaffCC \cite{javadiabhari_scaffccframeworkcompilation_2014} tracks qubit entanglement via data-flow analysis, and abstract interpretation techniques support assertion-based checking. These methods handle well-defined syntactic and structural defects effectively but cannot detect semantic bugs in which programs are syntactically valid yet algorithmically incorrect. Dynamic approaches complement static analysis by injecting faults. Quito \cite{wang_quito_2021} and QuSBT \cite{wang_qusbt_2022} produce quantum-specific test suites through coverage criteria and search-based testing, respectively. Mutation analysis tools such as QMutPy \cite{fortunato_qmutpy_2022} and Muskit \cite{mendiluze_muskit_2022} inject faults (gate additions, deletions, parameter perturbations) to assess test suite adequacy. However, these tools evaluate test effectiveness rather than performing end-to-end debugging, and none incorporate LLM-based reasoning. At a more foundational level, formal verification methods have been extended to the quantum domain. Quantum Hoare Logic (QHL) and its applied variant (aQHL) \cite{zhou_applied_2019} enable formal correctness proofs, while projection-based runtime assertions support dynamic state checking without collapsing valid quantum states. Although rigorous, these approaches require manually specified correctness properties and are difficult to scale to the diverse, rapidly evolving codebases of modern quantum frameworks.

Beyond bug detection, APR techniques aim to generate patches for buggy quantum code. Synthesis-based tools such as UnitAR \cite{li_automatic_2024} and HornBro \cite{tan_hornbro_2025} generate replacement unitary operations to restore correct behaviour, but they are limited to gate-level defects and often increase circuit depth. LLM-based repair offers more flexibility. For example, Guo et al. \cite{guo_repairing_2024} show that ChatGPT fixed 29 of 38 Bugs4Q bugs, but only with multi-round, human-provided hints and error descriptions. Recent work \cite{yoshida_leveraging_2026} augments GPT-5 prompts with QMutPy mutation analysis results and stack traces, achieving 94.4\% repair success on a Bugs4Q subset. However, this pipeline depends on external mutation analysis as a prerequisite, evaluates only single-turn repair, and is restricted to SDK-specific programs.

Our QBugLM addresses these limitations with a self-contained, end-to-end agentic framework that unifies bug generation, detection, repair, and validation. It differs from prior work in four key respects: (1) it operates on quantum SDK-agnostic OpenQASM 3.0 programs; (2) it employs a multi-agent LLM-based architecture with configurable prompting strategies and iterative feedback loops; (3) it can generate evaluation data automatically via a configurable, taxonomy-driven mutation generator (QBugGen), removing the reliance on pre-existing benchmarks; and (4) it validates repairs using simulation-based total variation distance, enabling fully automated, human-out-of-the-loop evaluation. 

\section{Conclusions and Future Work}
In this paper, we presented QBugLM, a multi-agent benchmarking framework for systematically evaluating LLM-based quantum software bug detection and repair. The framework integrates four components, including taxonomy-driven mutation generation, LLM-based bug detection and repair agents, and simulation-based validation, forming a fully automated, end-to-end pipeline that operates on framework-agnostic OpenQASM 3.0 programs. Through a comprehensive case study benchmarking two state-of-the-art LLMs, Claude 4.6 Sonnet and Qwen3 Coder Next, across different prompting strategies, bug categories, and quantum circuits. Our results also reveal three key findings, including (1) iterative feedback is critical, with a single retry raising Pass@1 from below 25\% to above 80\% for both LLMs, (2) simpler prompting strategies can outperform more elaborate alternatives (CoT, ReAct) for reasoning-capable models under fixed resource constraints; and (3) the open-source Qwen3 Coder Next achieves comparable accuracy to the proprietary Claude 4.6 Sonnet at substantially lower cost, though performance gaps widen on semantically more complex quantum circuits.

Several directions remain for future work. 
First, we plan to expand the mutation corpus by including multiple mutants per operator and incorporating multi-fault injection to improve statistical diversity and better approximate real-world bug distributions. Second, we aim to scale the evaluation to larger circuits and additional quantum frameworks to assess the generalisability of our findings across circuit complexities and programming paradigms. Third, we will explore hybrid agent configurations in which different models serve as the finder and fixer agents, and investigate techniques such as Retrieval-Augmented Generation and the use of framework documentation to improve repair accuracy for the most challenging semantic bugs. Furthermore, we will compare the LLM-based approach with other techniques to identify when agentic methods outperform static, dynamic, and formal analyses. Finally, we will extend the framework to handle compound bugs and assess regression detection, where repairs introduce new defects, better reflecting iterative quantum software development.

% \section*{Data Availability}
% The data and experimental results supporting the findings of this study are provided as anonymised supplementary artifacts within the submission system for the purpose of peer review.

\bibliographystyle{ieeetr}
\bibliography{references}

\end{document}